\documentclass[a4paper]{mem}
\usepackage{natbib}
\usepackage{amssymb}
\usepackage{graphicx}
\usepackage[a4paper]{hyperref}
\begin{document}
   \title{Population synthesis of $s$-process element enhanced stars:
     Constraining the $^{13}$C efficiency
}

   \author{A. Bona\v{c}i\'c Marinovi\'c,
     R.G. Izzard, M. Lugaro
        \and
           O.R. Pols
}

%   \offprints{P.G. Prada Moroni}
%\mail{via Dodecaneso 33, 16146 Genova }

   \institute{Sterrenkundig Instituut, Universiteit Utrecht, P.O. Box 80000, NL-3508 TA Utrecht, The Netherlands.\\\email{bonacic@astro.uu.nl}%\\ 
             }

   \abstract{
     We study $s$-process element abundance ratios in stars by carrying
     out stellar population synthesis, using a rapid synthetic stellar
     evolution code which includes an up-to-date treatment of
     AGB nucleosynthesis and evolution.
     In contrast to other studies, we find that a large spread in
     the $^{13}$C efficiency parameter ($^{13}{\rm C}_{\rm eff}$) is not
     needed to explain the observed spread in the ratios of heavy
     $s$-process to light $s$-process elements
     ([hs/ls]), but this comes naturally from the range of different
     initial stellar masses and their time evolution. As a result, the
     $^{13}$C efficiency needed for fitting most stars in the galactic
     disk is constrained to $1\lesssim ^{13}{\rm C}_{\rm eff}\lesssim 2$.
     In the same fashion we
     also study the [Pb/Ce] ratios of lead stars and find out that for
     low metallicities $^{13}{\rm C}_{\rm eff}\sim 0.5$.

%   \keywords{CM diagram --
%                He burning stars --
%                Isochrone fitting
%               }
   }
   \authorrunning{A. Bona\v{c}i\'c M. et al.}
   \titlerunning{Constraining $^{13}{\rm C}_{\rm eff}$ with population synthesis}
   \maketitle
%
%________________________________________________________________

\section{Introduction}

$S$-enhanced stars show high
abundances of elements heavier than Fe compared to the Sun,
which are produced via slow
neutron capture processes ($s$-process elements). These $s$-process 
elements are synthesized during the thermally pulsing asymptotic
giant branch (TP-AGB) phase in low and intermediate mass stars, however,
there are also less evolved stars which show over-abundances of these
elements, thus $s$-enhanced stars are classified as intrinsic and
extrinsic. Intrinsic $s$-enhanced stars (TP-AGB or post-AGB stars)
produce their own $s$-process elements including the radioactive element
Tc, which is observed in their envelopes would have already decayed
if it was not produced
in situ. These stars are factories of $s$-process
elements in the universe so understanding them is crucial to the
understanding of the chemical evolution of heavy elements in galaxies.
On the other hand, extrinsic $s$-enhanced stars are not
yet evolved enough to be in the TP-AGB phase, but have
accreted mass enriched with $s$-process elements from an intrinsic
$s$-enhanced companion in a binary system. Thus they show an overabundance
of stable
$s$-process elements, but no signatures of Tc.
The study of these stars is useful for probing the
nucleosynthesis that occurred in their WD companions when they were
on the TP-AGB phase, and it also provides information about
stellar interaction in wide binary systems, e.g., on the different
modes of mass transfer and tidal interaction.

Detailed stellar evolution and nucleosynthesis models must be able to
reproduce the abundances
of both intrinsic and extrinsic $s$-enhanced stars. 
Testing the absolute elemental 
abundances directly turns out to be difficult
because they depend on the dilution of material from the inter-shell
into the envelope, which in turn depends on several uncertain features
such as the amount of dredge up, mass loss, mass accretion, etc.
However, the
$s$-process element abundance ratios are practically unaffected
by these processes. They provide useful constraints on another
important uncertain parameter: the total neutron flux. This is determined
in the models of Gallino et al. (1998) by setting the amount of
$^{13}$C nuclei ($^{13}$C efficiency, i.e., $^{13}{\rm C}_{\rm eff}$),
which act as the main neutron source  
via the $^{13}{\rm C}(\alpha,n)^{16}{\rm O}$ reaction during the
inter-pulse periods.
Gallino et al. (1998) also introduce a $^{13}{\rm C}_{\rm eff}$
scale where the
standard case (ST in their paper, i.e., $^{13}{\rm C}_{\rm eff}=1$)
corresponds to the amount of $^{13}$C needed for
a 1.5 $M_{\odot}$ AGB star with half solar metallicity to
produce the main $s$-process element abundance of the Sun.
It was subsequently found that a wide spread in $^{13}{\rm C}_{\rm eff}$
(around two orders of magnitude) is needed for
the detailed models to reproduce the spread in abundance ratios
from the observational data (e.g., Busso et al., 2001, Gallino
et al., 2005, and references therein). However, for
computing time reasons, these authors used only a limited number of
initial masses for their models and only compared to the observations the
abundances for a ``typical'' thermal pulse. By carrying out
stellar population synthesis with a rapid synthetic code,
we can make models for a much finer grid of initial masses
and also take into account the time evolution of the abundances.
We can  thus produce a much larger set of synthetic data, and
weight the results with the initial mass function (IMF), in order
to compare to observations in a statistical sense. Thus we are able
to find better constraints on the $^{13}{\rm C}$ efficiency parameter.

%we are
%able to study better how accurate the detailed models are also to
%find better constrains on them.
%We can make a much finer grid of initial masses which providing a larger
%number of synthetic data to statistically compare with observations,
%such as weighting by the initial mass function (IMF) for stars and
%taking into account the change in time of the abundances.

%-------------------------------------------------------------
   \begin{figure*}
   \centering
   \includegraphics[width=14cm,height=12cm]{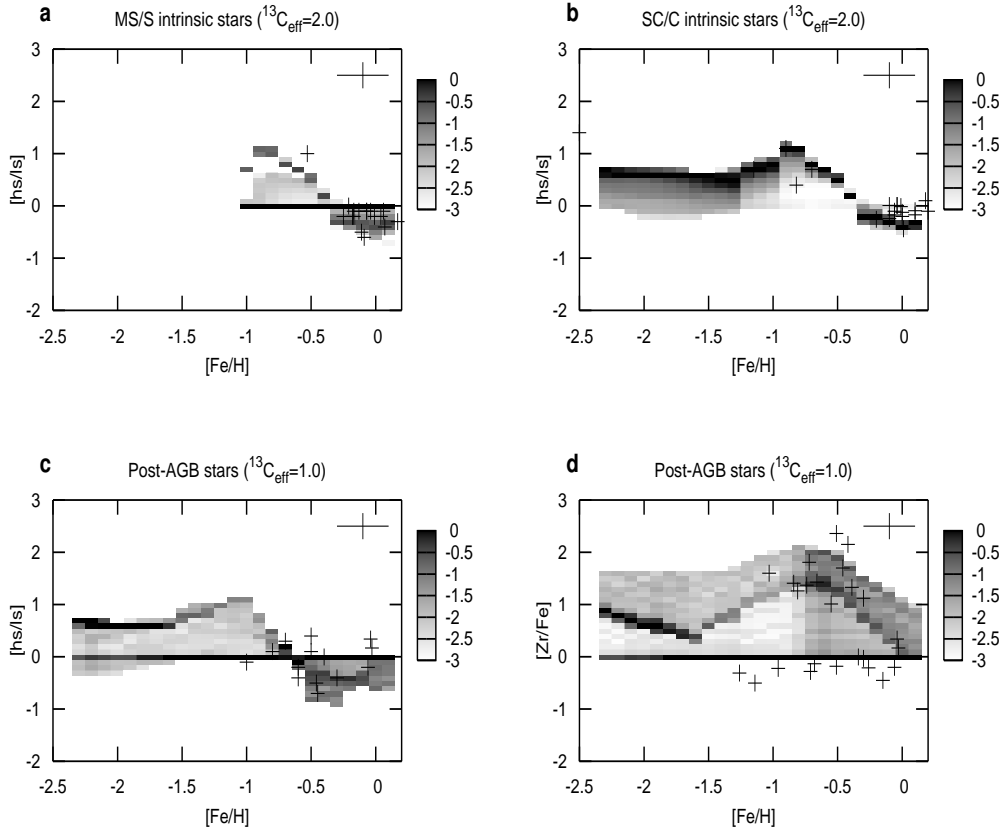}
      \caption{Intrinsic $s$-enhanced stellar population synthesis
	results	compared against the observations.  
	The gray scale is a logarithmic
	measure of the number distribution of stars over
	[hs/ls] or (in panel d) [Zr/Fe]. The crosses are the
	observational data (see the references in
	the text), which have an average
	error given by the size of the upper right cross in each plot.
	The number density is weighted by the IMF and by the time
	each star
	spends in an abundance bin, and then normalized for each
	metallicity.
      }
         \label{model_comp}
   \end{figure*}

%_____________________________________________________________
%

%-------------------------------------------------------------
   \begin{figure*}
   \centering
   \includegraphics[width=14cm,height=12cm]{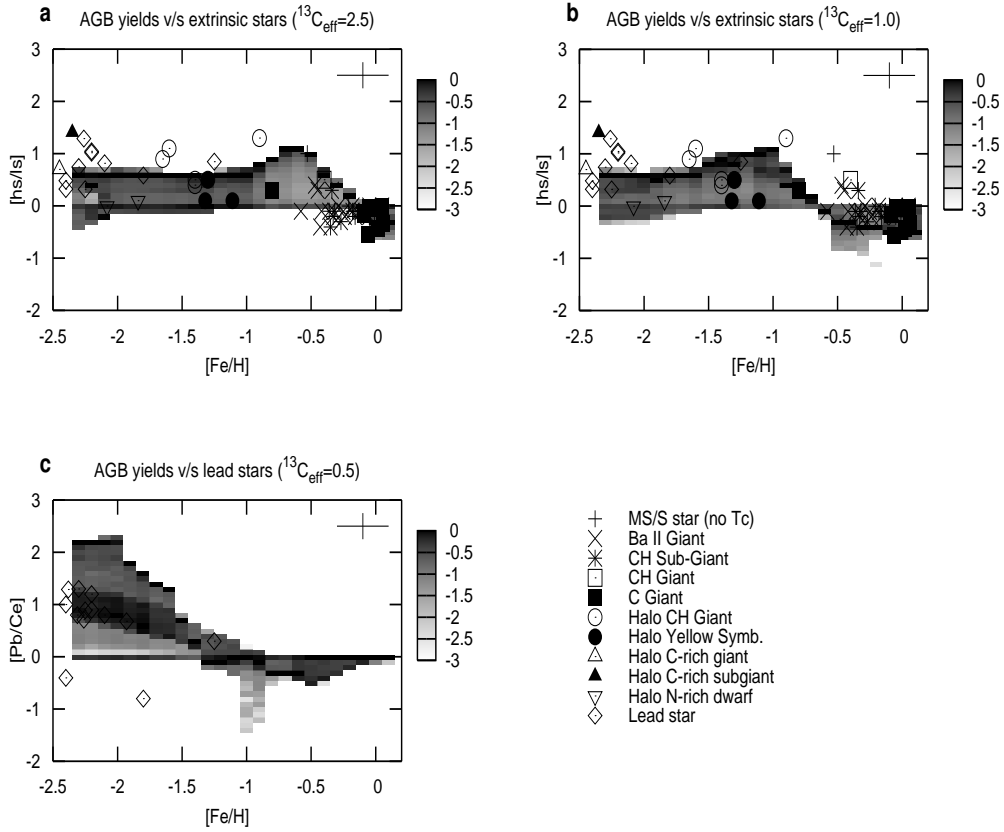}
      \caption{Population synthesis yield results
	compared against observations of extrinsic $s$-enhanced stars.  
	The gray scale is a logarithmic distribution of the
	yield ratios ([hs/ls] and [Pb/Ce]) for a population of stars
	as a function of metallicity.
	The points are observed data from different
	references (see text), which have an average
	error given by the size of the upper right cross in each plot.
	The results are weighted by the IMF and normalized for each
	metallicity.
              }
         \label{model_comp}
   \end{figure*}

%_____________________________________________________________
%

%-------------------------------------------------------------
   \begin{figure*}
   \centering
   \includegraphics[width=13cm]{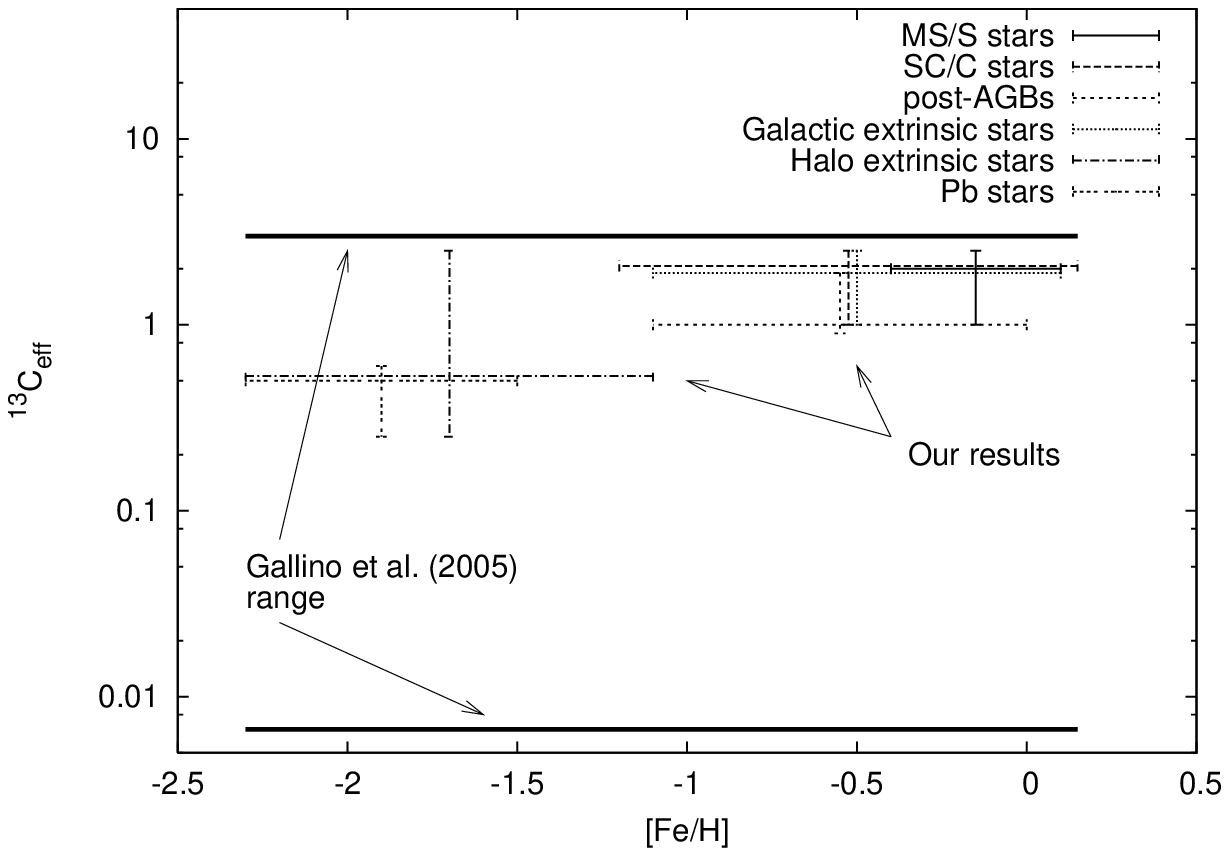}
   \caption{Range of $^{13}{\rm C}_{\rm eff}$ for which
        the observations of different kinds of
	$s$-enhanced stars are reproduced by population synthesis
	and the metallicity range within which
	they are observed. The thick solid lines show the range
	of $^{13}{\rm C}_{\rm eff}$ needed by Gallino et al. (2005)
	to reproduce the observations.
           }
         \label{model_comp}
   \end{figure*}

%_____________________________________________________________
%

\section{Population synthesis}

We carry out the population synthesis with a rapid synthetic
evolution code based on that from Izzard et al. (2004), but
modified to be more self-consistent
with the evolutionary phases prior to the TP-AGB
(Bonacic et al. in preparation). The $s$-process nucleosynthesis
is calculated by interpolating the results of Gallino et al.
(1998), using a grid of inter-shell abundances for different
values of stellar mass, metallicity, $^{13}{\rm C_{eff}}$ and
thermal pulse number.
We investigate a fine grid of initial stellar
masses and we follow the time evolution of the surface
abundances in each star. We explore the ratio of the heavy
$s$-process element (hs) average abundance
to the light $s$-process element (ls) average abundance
\begin{eqnarray*}
\rm{
\left[\frac{hs}{ls}\right]=\left(\left[\frac{Ba}{Fe}\right]+
\left[\frac{La}{Fe}\right]+\left[\frac{Ce}{Fe}\right]+
\left[\frac{Nd}{Fe}\right]+\right.}\\
\rm{
\left.\left[\frac{Sm}{Fe}\right]\right)*0.2-
\left(\left[\frac{Y}{Fe}\right]+\left[\frac{Zr}{Fe}\right]\right)*0.5}
\end{eqnarray*}
in 100 TP-AGB stars for each metallicity in a range of initial masses from
1.0 to 8.0 M$_{\odot}$. We have taken into account 25 different
metallicities in the range $-2.3<$[Fe/H]$<0.1$ and for each one
we have weighted the output of the different masses by the IMF of
Kroupa, Tout \& Gilmore (1993).

The population synthesis is carried out for different values of
$^{13}{\rm C}_{\rm eff}$ and the output is compared to observational
data, as described below.

%Stellar wind mass loss on the AGB as described by \citet{VW93}.

\section{Comparison of results with observations}
\label{sec3}

\subsection{Intrinsic $S$-enhanced stars}

In the context of single stellar evolution it
is straightforward to compare the output of the population
synthesis with the observed intrinsic $s$-enhanced stars.
The first set of data that we compare to are the intrinsic MS/S stars
of Busso et al. (2001),
for which we select from the synthesized population the
stars that have a surface effective temperature less than 3500 K
and that are not carbon stars, i.e., C/O$<0.95$. As can be
seen in Fig. 1a, the observations can be fitted quite well
with just one $^{13}C$ efficiency value, $^{13}{\rm C}_{\rm eff}=2$.

The next set are intrinsic SC and C stars, taken from
Busso et al. (2001) and from Abia et al. (2002). Our synthesized
SC and C stars are those TP-AGB stars which have C/O$>0.95$.
Again, from Fig. 1b it can be seen that with only $^{13}{\rm C}_{\rm eff}=2$
the data is well fitted, without the need for a spread in
$^{13}{\rm C}_{\rm eff}$.

The last set of intrinsic stars to compare to are the post-AGB
stars, taken from Busso et al. (2001), Van Winckel (2003),
Giridhar \& Arellano (2005) and Gallino et al. (2005).
We have considered as post-AGB stars all those from our synthetic
TP-AGB sample which have an envelope mass of 0.02 $M_{\odot}$
or less. Most of them can be well fitted with $^{13}{\rm C}_{\rm eff}\approx 1$
(Fig. 1c), with a few exceptions which need
$^{13}{\rm C}_{\rm eff}\sim 1.5$. It
is important also to notice that there is an
apparent split in the observed [Zr/Fe], which suggests that some
post-AGB stars (those with [Zr/Fe]$\approx 0$) did not experience dredge-up
episodes, while others (those highly $s$-process enhanced) did experience
dredge-up (Van Winckel,
2003). This dichotomy is also observed in our results and 
is consistent with $^{13}{\rm C}_{\rm eff}\approx 1$ for these stars,
as can be seen in Fig. 1d.

\subsection{Extrinsic $S$-enhanced stars}

Extrinsic $s$-enhanced stars can also be studied with single stellar
population synthesis albeit only in a an approximate way.
As these stars are formed by accreting mass from
an $s$-enhanced companion, their abundance enhancements are (to first
order) proportional to the mass of the various elements yielded by
their companions. We thus compare to the yields, rather than to the
surface abundances, of our synthesized TP-AGB stars, again
weighted by the IMF.
Comparing the population synthesis yields with the
extrinsic star data of Busso et al. (2001) and Abia et al. (2002)
we see in Figs. 2a and 2b that the [hs/ls] ratios of
these stars are well reproduced with
$1\lesssim ~^{13}{\rm C}_{\rm eff}\lesssim 3$, which is consistent with
the values found for the intrinsic $s$-enhanced stars.

At [Fe/H]$\lesssim -1$ there are no intrinsic stars nor is the [hs/ls]
ratio very sensitive to $^{13}{\rm C}_{\rm eff}$ changes.
It is at this point that
the role of lead stars becomes important. They are low
metallicity $s$-enhanced
extrinsic stars on which Pb has been detected. The [Pb/hs] ratios are
sensitive to the value of $^{13}{\rm C}_{\rm eff}$. In
particular we compute the [Pb/Ce] ratios and compare our models to the
data of Van Eck et al. (2003) and references therein. We find
that for most of the lead star [Pb/Ce] ratios can be reproduced with
$^{13}{\rm C}_{\rm eff}\approx 0.5$ (see Fig. 2c), except for two
clear outliers.

\section{Conclusions}   

We have found that in order to reproduce the different [hs/ls] and
[Pb/Ce] ratios observed in $s$-enhanced stars only a small spread in the
$^{13}$C efficiency is needed. The observed spread in element ratios
can be naturally
explained by different initial stellar masses and the time evolution
of the TP-AGB stars, and it is not necessary to use the large spread
of $^{13}{\rm C}_{\rm eff}$ values (more than two orders of magnitude)
suggested by other works, e.g., Busso et al., 2001, Gallino
et al., 2005, and references therein.
Galactic disk objects ([Fe/H]$\gtrsim -1$) are well reproduced
by considering the range $1\lesssim ~^{13}{\rm C}_{\rm eff}\lesssim 3$ in our
population synthesis. Most halo objects (low-metallicity stars) 
are well reproduced with $^{13}{\rm C}_{\rm eff}\approx 0.5$. This might
suggest that at low metallicities $^{13}{\rm C}_{\rm eff}$ is smaller.
Fig. 3 shows a summary of the 
range of $^{13}{\rm C}_{\rm eff}$ values needed to reproduce the different
types of stellar population, which is limited within a factor of 5.

In the results presented we did not take into account the existence
of an age-metallicity relation, which limits the range of stellar masses
that actually contributes to the observed intrinsic $s$-enhanced stars
at each metallicity.In future work we will take this into account, and
also perform a proper binary evolution population synthesis to get a
more reliable outcome for the extrinsic $s$-enhanced stars.

%\begin{acknowledgements}
%\end{acknowledgements}

\bibliographystyle{aa}

\end{document}